\documentclass[12pt]{article}

\ifx\pdfoutput\undefined
\usepackage[dvips,bookmarks]{hyperref}
\else
\usepackage{hyperref}
\fi
\hypersetup{colorlinks=false,bookmarksopen,bookmarksnumbered,citecolor=blue,
   pdfstartview=FitH}

\usepackage{latexsym}
\usepackage{amssymb,amsfonts,amsmath}
\usepackage{graphicx} 
\usepackage{indentfirst}
\usepackage{bbm}

\parskip=\medskipamount

\newcommand {\cA}{{\cal A}}
\newcommand {\cB}{{\cal B}}
\newcommand {\cC}{{\cal C}}
\newcommand {\cD}{{\cal D}}

\newcommand {\cJ}{{\cal J}}

\newcommand {\cL}{{\cal L}}

\newcommand {\cN}{{\cal N}}

\newcommand {\cT}{{\cal T}}

\def\a{\alpha}
\def\b{\beta}

\def\d{\delta}
\def\e{\epsilon}
\def\f{\phi}
\def\g{\gamma}
\def\G{\Gamma}

\def\j{\psi}
\def\k{\kappa}
\def\l{\lambda}
\def\m{\mu}
\def\n{\nu}
\def\o{\omega}

\def\q{\theta}
\def\r{\rho}
\def\s{\sigma}
\def\t{\tau}

\def\D{\Delta}
\def\F{\Phi}
\def\J{\Psi}
\def\L{\Lambda}
\def\O{\Omega}

\def\S{\Sigma}
\def\U{\Upsilon}
\def\X{\Xi}

\def \bi{\bibitem}

\def\ri{{\rm i}}
\def\re{{\rm e}}

\newcommand{\ve}{\varepsilon}                            

\newcommand{\hf}{\frac12}

\newcommand{\vf}{\varphi}

\newcommand{\be}{\begin{equation}}
\newcommand{\ee}{\end{equation}}
\newcommand{\bea}{\begin{eqnarray}}
\newcommand{\eea}{\end{eqnarray}}
\newcommand{\non}{\nonumber}
\newcommand{\ba}{\begin{array}}
\newcommand{\ea}{\end{array}}

\newcommand{\au}{\underline{a}}

\newcommand{\iu}{\underline{i}}
\newcommand{\ju}{\underline{j}}

\def\dt#1{{\buildrel {\hbox{\LARGE .}} \over {#1}}}    

\newcommand{\bm}[1]{\mbox{\boldmath$#1$}}

\def\double #1{#1{\hbox{\kern-2pt $#1$}}}

\newcommand{\ha}{{\hat{a}}}
\newcommand{\hb}{{\hat{b}}}
\newcommand{\hc}{{\hat{c}}}
\newcommand{\hd}{{\hat{d}}}

\newcommand{\hal}{{\hat{\a}}}
\newcommand{\hbe}{{\hat{\b}}}
\newcommand{\hga}{{\hat{\g}}}
\newcommand{\hde}{{\hat{\d}}}

\newcommand{\alu}{{\underline{\a}}}
\newcommand{\beu}{{\underline{\b}}}
\newcommand{\gau}{{\underline{\g}}}

\newcommand{\sSp}{\mathsf{Sp}}
\newcommand{\sSU}{\mathsf{SU}}
\newcommand{\sSL}{\mathsf{SL}}
\newcommand{\sGL}{\mathsf{GL}}
\newcommand{\sSO}{\mathsf{SO}}
\newcommand{\sU}{\mathsf{U}}

\newcommand{\sOSp}{\mathsf{OSp}}

\newcommand{\sMat}{\mathsf{Mat}}
\newcommand{\sO}{\mathsf{O}}

\newcommand{\bsubeq}{\begin{subequations}}
\newcommand{\esubeq}{\end{subequations}}

\newcommand{\rd}{\mathrm d}
\begin{document}

\begin{titlepage}
\begin{flushright}
June 2012\\
Revised version: August 2012
\end{flushright}
\vspace{5mm}

\begin{center}
{\Large \bf Conformally compactified Minkowski superspaces revisited}
\\ 
\end{center}

\begin{center}

{\bf Sergei M. Kuzenko}

\footnotesize{
{\it School of Physics M013, The University of Western Australia\\
35 Stirling Highway, Crawley W.A. 6009, Australia}}  ~\\
\texttt{sergei.kuzenko@uwa.edu.au}\\
\vspace{2mm}

\end{center}
\vspace{5mm}

\begin{abstract}
\baselineskip=14pt
Starting from the standard supertwistor realizations for conformally compactified $\cN$-extended 
Minkowski superspaces in three and four space-time dimensions, we elaborate on 
alternative realizations in terms of graded two-forms on the dual supertwistor spaces. 
The construction is further generalized to the cases of 4D $\cN=2$ 
and 3D $\cN$-extended harmonic/projective superspaces.
We present  a superconformal  Fourier expansion of 
tensor superfields on the 4D $\cN=2$  harmonic/projective superspace. 
\end{abstract}
\vspace{0.5cm}

\vfill
\end{titlepage}

\newpage
\renewcommand{\thefootnote}{\arabic{footnote}}
\setcounter{footnote}{0}

 \tableofcontents


\numberwithin{equation}{section}




\section{Introduction}
Two years ago, Steven Weinberg \cite{Weinberg} argued  that the use of Dirac's  
projective lightcone realization
for compactified four-dimensional (4D) Minkowski space \cite{Dirac} greatly simplifies the calculation of Green's functions in conformally-invariant field theories 
(see also \cite{CPPR}). 
Clearly, this was not the first paper in which ($d+2$)-dimensional methods were used to study conformal theories in $d$ dimensions, see, e.g.,  
\cite{Kastrup,MackSalam,FGG,Ferrara,TMP,Todorov,Siegel,KY}. 
Nevertheless, Weinberg's work has stimulated 
some renewed interest in Dirac's construction \cite{Dirac} and its implications. 
In particular, Ref. \cite{GSS}  considered  a 4D  $\cN=1$  supersymmetric extension 
of the  projective lightcone formalism of \cite{Dirac}.  
A generalization of \cite{GSS} to the case  $\cN>1$ was attempted in \cite{Maio}.
Specifically, Refs. \cite{GSS,Maio} suggested to describe conformally compactified $\cN$-extended 
Minkowski superspace, $\overline{\mathbb M}{}^{4|4\cN}$,
in terms of graded two-forms on the dual supertwistor space.
Here we demonstrate how to derive such a description starting from 
the standard supertwistor realization for $\overline{\mathbb M}{}^{4|4\cN}$, see 
\cite{K-compact} and references therein. Our discussion is more complete and differs in some details. 

We also present a new realization for compactified 
4D $\cN=2$ harmonic/projective superspace building on the formulation given in  \cite{K-compact}. 
Finally, we generalize  our construction to the case 
of conformally compactified 3D $\cN$-extended Minkowski superspace $\overline{\mathbb M}{}^{3|2\cN}$ 
and its harmonic/projective extensions \cite{KPT-MvU} (see \cite{Howe:1994ms} for an alternative construction).

This paper is organized as follows. Section 2 is  devoted to non-supersymmetric warm-up exercises. Here we describe three different realizations for conformally compactified Minkowski space in four dimensions, 
$\overline{\mathbb M}{}^{4}$, and prove their equivalence.  
In section 3 we start by recalling the standard supertwistor realization for $\overline{\mathbb M}{}^{4|4\cN}$.
Then we introduce a novel bi-supertwistor realization for $\overline{\mathbb M}{}^{4|4\cN}$. 
After that we prove the equivalence of these two realizations. 
Section 4 is devoted to a new realization for compactified 
4D $\cN=2$ harmonic/projective superspace, $\overline{\mathbb M}{}^{4|8} \times {\mathbb C}P^1$. 
In sections 5 to 7, we generalize the construction to three space-time dimensions. 
Concluding comments are given in section 8.
The main body of the paper is accompanied by a technical appendix devoted to 
spinors in 4 + 2 dimensions.

\section{Compactified 4D Minkowski space} 
In this section we describe three different realizations for conformally compactified 
Minkowski space in four dimensions. 

\subsection{Dirac's realization in $d$ space-time dimensions} 
We start by recalling the  projective light cone formalism of \cite{Dirac}
(see also Weyl's book  \cite{Weyl}).
Consider a flat space ${\mathbb R}^{d,2} $
parametrized by Cartesian coordinates  
\bea
X^{\ha} =( X^a ,X^{d+1}, X^{d+2})~, \qquad a = 0,1,\dots, d-1
\eea 
and endowed with the metic $\eta_{\ha \hb} = {\rm diag}  (-1, +1,\dots, +1,-1)$.
Let us inroduce the cone $\cC$ in  ${\mathbb R}^{d,2} $ defined by 
\bea
\eta_{\ha \hb} X^{\ha} X^{\hb} = 0~. 
\label{cone}
\eea
The space of all straight lines belonging to  $\cC$ and passing through the origin of
 $ {\mathbb R}^{d,2}$ is known as Dirac's conformal space \cite{Dirac}
 or compactified Minkowski space, $\overline{\mathbb M}{}^d$. 
 It can be defined as the quotient space of $\cC - \{ 0 \} $ with respect to  the equivalence relation 
\bea
X^{\ha} ~\sim ~ \l \, X^{\ha} ~, \qquad \l \in {\mathbb R} -\{ 0 \}
\label{2.3}
\eea
which identifies all points on a straight line in $ {\mathbb R}^{d,2}$.
The conformal group in $d$ dimensions, $\sO (d, 2)/{\mathbb Z}_2$, 
with ${\mathbb Z}_2 = \{ \pm {\mathbbm 1}_{d+2} \}$, 
naturally acts on 
 $\overline{\mathbb M}{}^d$. It may be seen that  $\overline{\mathbb M}{}^d$ is a homogeneous 
 space of the connected conformal group, which is  $\sSO_0 (d, 2)$ if $d$ is odd and 
  $\sSO_0 (d, 2)/{\mathbb Z}_2$ if $d$ is even.

It follows that the global structure of   $\overline{\mathbb M}{}^d$ as a topological space  is
\bea
\overline{\mathbb M}{}^d = (S^{d-1} \times S^1)/{\mathbb Z}_2~.
\label{2.4}
\eea
Indeed, the constraint \eqref{cone} and the `gauge' freedom \eqref{2.3}
can be used to choose $X^{\au}$ such that 
\bea
(X^0)^2 + (X^{d+2})^2 = \sum_{i=1}^{d-1} (X^i)^2  + (X^{d+1})^2= 1~.
\eea
For such a choice, the equivalence relation \eqref{2.3} still allows us to identify 
$X^{\ha}$ and $-X^{\ha}$, which is the reason for  ${\mathbb Z}_2$  in \eqref{2.4}.
In four dimensions,  $\overline{\mathbb M}{}^4$ is the same 
topological space as the group manifold U(2). In three dimensions, 
$\overline{\mathbb M}{}^3$ can be identified with $\sU(2)/\sO (2)$.

Minkowski space ${\mathbb M}^d \equiv {\mathbb R}^{d-1,1}$ can be identified, e.g.,  
with the open {\it dense} domain of  $\overline{\mathbb M}{}^d$ 
on which $X^{d+1}  + X^{d+2} \neq 0$.\footnote{The closed subset  
$C_3:= \overline{\mathbb M}{}^d - {\mathbb M}{}^d$
can be identified with a lightcone in ${\mathbb R}^{d-1,1}$. Indeed, the points of $C_3$ can be uniquely parametrized by $(D+2)$-vectors of the form $X^{\ha} = (z^a, -\hf, \hf)$, where $z^a \in {\mathbb R}^{d-1,1}$ 
is null, $z^a z_a=0$. }
This domain can be parametrized by  variables 
\bea
x^a = \frac{X^a}{X^{d+1}  + X^{d+2}}
\eea
which are invariant under the identification \eqref{2.3}.
In terms of these coordinates, one obtains a standard action of the conformal group in  ${\mathbb M}^d$.
Thus $x^a$ can be identified with Cartesian coordinates for Minkowski space. 

In the remainder of this section our consideration is restricted to the case $d=4$. 

\subsection{Twistor realization} 
Here we recall the so-called twistor realization\footnote{This realization has been known since the 1960s,
see 
\cite{Uhlmann,Penrose,Segal,Todorov,WW} and references therein.} 
of compactified Minkowski space  $\overline{\mathbb M}{}^4$ 
as the set of null two-dimensional subspaces in the  twistor space, 
${\mathbb C}^4$, 
equipped with the inner product
\bea
\langle T, S \rangle = T^\dagger \,\O \, S~, \qquad
\O =\left(
\begin{array}{cc}
 0  &  {\mathbbm 1}_2\\
 {\mathbbm 1}_2 &      0
\end{array}
\right) ~,
\label{2.6}
\eea
for any twistors $T,S \in {\mathbb C}^4$. 
The components of a twistor 
$T$ and its dual  $\bar T := T^\dagger \O $ are denoted as 
\bea
T= (T_{\hal} ) 
= \left(
\begin{array}{c}
 f_\a \\
\bar h^{\dt \a} 
\end{array}
\right)~, \qquad \bar T = (\bar T^{\hal} )= ( h^\a , \bar f_{\dt \a} )~.
\eea

By construction, the inner product \eqref{2.6} is invariant under the action
of the group $\sSU(2,2)$ which is the two to one covering of 
the group $\sSO_0(4,2)$,  which  in turn is the two to one covering of
 the connected conformal group,  $\sSO_0 (4, 2)/{\mathbb Z}_2$.
The elements of $\sSU(2,2)$ will be represented by block matrices
\bea
g=(g_{\hal \,}{}^{\hbe} )= \left(
\begin{array}{cc}
 A  & \ri B\\
-\ri C &    D 
\end{array}
\right) \in {\sSL}(4,{\mathbb C}) ~, \qquad 
g^\dagger \,\O \,g = \O~,
\label{SU(2,2)}
\eea
where  $A,B,C$ and $D$ are $2\times 2$ matrices.
Under the action of $\sSU(2,2)$, the twistor $T_{\hal}$ and its dual $\bar T^{\hal}$ transform
as follows:
\bea
T_{\hal} ~\to ~g_{\hal \,}{}^{\hbe} T_{\hbe}~, \qquad 
\bar T^{\hal} ~\to ~ \bar T^{\hbe} (g^{-1})_{\hbe \,} {}^{\hal}~.
\eea
These two representations are inequivalent.  

We denote by  $\overline{\mathbb M}{}^4_{\text{T}}$ the space of null two-planes through 
the origin in ${\mathbb C}^4$. Given such a two-plane, it is generated by two linearly independent 
twistors, $T^\m = ( T^1, T^2 )$, subject to  the null conditions
\bea
\langle T^\m, T^\n \rangle \equiv \bar T^\m T^\n= 0~, \qquad 
\m ,\n =1,2~.
\label{nullplane11}
\eea 
The basis  chosen, $\{T^\m \}$, is defined only modulo 
the equivalence relation 
\bea
\{ T^\m \}~ \sim ~ \{ \tilde{T}^\m \} ~, \qquad
\tilde{T}^\m = T^\n\,C_\n{}^\m~, 
\qquad C \in {\sGL}(2,{\mathbb C}) ~.
\label{nullplane22}
\eea
The two bases, $\{T^\m \}$ and $\{ \tilde{T}^\m \}$, 
define one and the same two-plane in ${\mathbb C}^4$.

Minkowski space, ${\mathbb M}^4 \equiv {\mathbb R}^{3,1}$,
can be embedded into $\overline{\mathbb M}{}^4_{\rm T}$ as a dense open subset. 
Consider the open domain of $\overline{\mathbb M}{}^4_{\rm T}$ consisting of those 
null two-planes which have the form
\bea
({ T}_{\hal}{\,}^\m)
= \left(
\begin{array}{c}
{ F}_\a{}^\m  \\
{ H}^{\dt \a \m}
\end{array}
\right) ~, \qquad \det ( { F}_\a{}^\m ) \neq 0~.
\eea
Then,  choosing $C= F^{-1}$ in \eqref{nullplane22} and making use of the null condition 
\eqref{nullplane11}
leads to the basis
\bea
({ T}_{\hal}{\,}^\m)= \left(
\begin{array}{r}
\d_\a{}^\b \\  -{\rm i}\, {x}^{\dt \a \b}
\end{array}
\right) 
~, \qquad  {x}^{\dt \a \b} :=  x^m (\tilde{\s}_m)^{\dt \a \b}~, \qquad x^m = (x^m)^*~.
\eea
The group element \eqref{SU(2,2)} acts on $\tilde{x} = ( {x}^{\dt \a \b}) $ by the fractional 
linear transformation
\bea
\tilde{x}' = { (C+ D \tilde{x} )} {(A+B \tilde{x})}^{-1}~,
\eea
which is a standard conformal transformation in Minkowski space ${\mathbb M}^4$.
Therefore the open domain of $\overline{\mathbb M}{}^4_{\rm T}$ introduced can indeed be identified with 
Minkowski space. 

Let us show that $\overline{\mathbb M}{}^4_{\text{T}}$ is equivalent 
to the compactified Minkowski space introduced in the previous subsection. 
Using the two twistors $T^\m$, which describe a point in 
$\overline{\mathbb M}{}^4_{\text{T}}$, 
we define the following $4\times 4$ matrix:
\bea
Y_{\hal \hbe} := T_{\hal}{}^\m T_{\hbe}{}^\n \ve_{\m \n}~, \qquad \ve_{\m \n }= -\ve_{\n \m}~, \qquad
\ve_{12} =-1~.
\label{2.166}
\eea
This matrix is antisymmetric, 
\bea
Y_{\hal \hbe} = - Y_{\hbe \hal} ~,
\eea
and defined modulo the equivalence relation 
\bea
Y_{\hal \hbe} ~\sim ~c \, Y_{\hal \hbe}~, \qquad c = \det (C_{\hal}{}^{\hbe} ) \in {\mathbb C}-\{0\}~.
\eea
It is characterized by the algebraic properties
\bea
Y_{[\hal \hbe} Y_{\hga ] \hde} =0 \quad \longleftrightarrow \quad
Y_{[\hal \hbe} Y_{\hga  \hde ]} =0 ~.
\label{2.15}
\eea
Using the dual twistors $\bar T$, we can define 
\bea
\bar Y^{\hal \hbe} := \ve_{\m \n } \bar T^{\m \, \hal} \bar T^{\n\, \hbe}~.
\eea
The matrices $Y= (Y_{\hal \hbe})$ and $\widetilde{\bar Y} = (\bar Y^{\hal \hbe})$ are related 
to each other as follows
\bea
\widetilde{\bar Y} = - \O Y^\dagger \O~.
\label{17conj}
\eea
As a consequence of \eqref{nullplane11}, we have 
\bea
\bar Y^{\hal \hga} Y_{\hga \hbe}=0~.
\label{2.17}
\eea

Introduce six-vectors $Y^{\ha}$ and $\bar Y^{\ha}$ defined by 
\bea
Y^{\ha} := \frac{1}{4} (\S^{\ha})_{\hbe \hga} Y^{\hbe \hga} ~, \qquad 
\bar Y^{\ha} := \frac{1}{4} ( \S^{\ha})_{\hbe \hga} \bar Y^{\hbe \hga} ~, 
\eea
with $ Y^{\hal \hbe} := \hf \ve^{\hal\hbe\hga \hde}  Y_{\hga \hde}$.
These six-vectors are mutually null, 
\bea
\eta_{\ha \hb} Y^{\ha} Y^{\hb} 
= \eta_{\ha \hb } \bar Y^{\ha} \bar Y^{\hb}
= \eta_{\ha \hb} Y^{\ha} \bar Y^{\hb} 
=0~.
\eea
The first two relations follow from \eqref{A.14}, while the third is a consequence of 
 \eqref{2.17}. Moreover, it follows from \eqref{2.17} that 
\bea 
Y^{\ha} \bar Y^{\hb} (\S_{\ha \hb} )_{\hga}{\,}^{\hde} =0~,
\eea
and therefore the mutually conjugate six-vectors $Y^{\ha}$ and 
$\bar Y^{\ha}= (Y^{\ha})^*$ are linearly dependent. 
This means that 
\bea
Y^{\ha} = \re^{\ri \vf } X^{\ha}~, \qquad \bar Y^{\ha} = \re^{-\ri \vf } X^{\ha}~,
\qquad \vf \in {\mathbb R}~,
\eea
for some real null six-vector $X^{\ha}$. By construction, $X^{\ha}$
is defined modulo the equivalence relation \eqref{2.3}. 
In summary, we have defined an injective\footnote{For completeness, we prove that the map $F$ is injective.
Suppose this is not true and there exist  two different two-planes that are mapped by $D$ to the same point of 
$\overline{\mathbb M}{}^4$. Then we can choose the bases $T^\m =(T,S) $ and $T'^\m = (T',S') $ for the two-planes under 
consideration such that $T\wedge S = T' \wedge S'$ (here we think of $Y_{\hal \hbe} $ defined by \eqref{2.166}
as an element of $\wedge^2{\mathbb C}^4$). Let us introduce a basis for ${\mathbb C}^4$ consisting of $T^\m$ and two additional vectors $W^\m$. 
Without loss of generality, we can choose $T'$ and $S'$ such that  
$T' = T + t_\m W^\m$ and $S' = S+s_\m W^\m$, for some complex parameters $t_\m$ and $s_\m$. 
Now, since $T^\m$ and $W^\m$ are linearly independent, the condition $T\wedge S = T' \wedge S'$ tells us that 
$t_\m = s_\m=0$. As a result, $T^\m  $ and $T'^\m  $ define one and the same two-plane, which contradicts 
the assumption.} 
map 
$F: \overline{\mathbb M}{}^4_{\text{T}} \to \overline{\mathbb M}{}^4$. 
This map is in fact onto, and therefore one-to-one. 
To prove this, we associate with $X^{\ha}$  the antisymmetric matrix
\bea
X_{\hal \hbe} := X^{\ha} (\S_{\ha})_{\hbe \hga} ~.
\eea
For its conjugate $\widetilde{\bar X} = (\bar X^{\hal \hbe} )$ 
defined according to \eqref{17conj} we obtain
\bea
\bar X^{\hal \hbe} = 
\hf \ve^{\hal\hbe\hga \hde}  X_{\hga \hde} \equiv  X^{\hal \hbe} ~.
\eea
The matrices $X_{\hal \hbe} $ and $\bar X^{\hal \hbe} $ obey the properties 
\eqref{2.15} and \eqref{2.17}. It turns out that $X_{\hal \hbe} $ defines a null two-plane in 
${\mathbb C}^4$. This follows from the discussion below. 

\subsection{Bi-twistor  realization} 
There exists  an alternative realization for  $\overline{\mathbb M}{}^4_{\text{T}}$. 
Let us denote by $\cL$ 
the set of all {\it non-zero} complex antisymmetric matrices $Y_{\hal \hbe} = - Y_{\hbe \hal}$ 
which obey the algebraic constraints 
\begin{subequations} 
\bea
Y_{[\hal \hbe} Y_{\hga  \hde ]} &=&0~, \label{2.25a} \\ 
\bar Y^{\hal \hga} Y_{\hga \hbe}&=&0~, \qquad \widetilde{\bar Y} := - \O Y^\dagger \O~.
\label{2.25b}
\eea
\end{subequations}
We introduce the quotient space $\overline{\mathbb M}{}^4_{\text{BT}} = \cL / \! \!\sim$, 
where the equivalence relation is defined by 
\bea
Y_{\hal \hbe} ~\sim ~c \, Y_{\hal \hbe}~, \qquad c  \in {\mathbb C}-\{0\}~.
\eea
We will show that $\overline{\mathbb M}{}^4_{\text{BT}}$ can be naturally identified with 
$\overline{\mathbb M}{}^4_{\text{T}}$.

It follows from \eqref{2.25a} that $Y_{\hal \hbe} $ is decomposable, that is 
\bea 
Y_{\hal \hbe} = T_{\hal} S_{\hbe}- T_{\hbe} S_{\hal}~, 
\eea
for some linearly independent twistors $T_{\hal}$ and $S_{\hal}$, 
see e.g. \cite{WW} for the proof. Now, eq. \eqref{2.25b} is equivalent to 
\bea
\bar Y^{\hal \hga} Y_{\hga \hbe} = \bar T^{\hal} \Big\{ 
\langle S, T \rangle S_{\hbe} - \langle S, S \rangle T_{\hbe} \Big\}
+  \bar S^{\hal} \Big\{ 
\langle T, S \rangle T_{\hbe} - \langle T, T \rangle S_{\hbe} \Big\} =0~.
\eea
Since $ \bar T^{\hal} $ and $ \bar S^{\hal} $ are linearly independent dual twistors, 
the expressions in figure brackets must vanish. Since  $T_{\hal}$ and $S_{\hal}$
are linearly independent, we conclude that 
\bea
\langle T, T \rangle = \langle S, T \rangle =\langle T, S \rangle = \langle S, S \rangle =0~, 
\eea
and therefore the two-plane in ${\mathbb C}^4$ associated with $T_{\hal}$ and $S_{\hal}$
is null. We finally choose $T^\m = (T, S)$.

\section{Compactified 4D Minkowski superspace} 

Supertwistor space ${\mathbb C}^{4|\cN}$ was introduced by Ferber \cite{Ferber} as a supersymmetric extension of the twistor space. 
The elements of  ${\mathbb C}^{4|\cN}$ are called supertwistors. We use capital boldface letters, 
${\bm T}, {\bm S}, \dots$, to  denote supertwistors, for instance 
\bea
{\bm T} = ({\bm T}_A) = \left(
\begin{array}{c}
{\bm T}_{\hal}  \\
{\bm T}_i 
\end{array}
\right) 
~, \qquad i = 1, \dots, \cN~.
\label{3.1}
\eea
The supertwistor space is equipped with the inner product 
\bea
\langle {\bm T}, {\bm S} \rangle = {\bm T}^\dagger \, {\bm \O} \, {\bm S}~, \qquad
{\bm \O} =
\left(
\begin{array}{ccc}
0&  {\mathbbm 1}_2 &0\\
{\mathbbm 1}_2 &0 &0\\
0 & 0& -{\mathbbm 1}_{\cN}
\end{array}
\right) = 
\left(
\begin{array}{cc}
\O &0\\
 0& -{\mathbbm 1}_{\cN}
\end{array}
\right) 
~.
\eea
This inner product is invariant under the $\cN$-extended superconformal group ${\sSU}(2,2|\cN) $
spanned by supermatrices of the form 
\bea
g = (g_A{}^B) \in {\sSL}(4|\cN ) ~, \qquad 
g^\dagger \, {\bm \O} \,g = {\bm \O}~.
\label{SU(2,2|N)}
\eea
Associated with a supertwistor $\bm T$, eq. \eqref{3.1}, is its dual 
\bea
\bar {\bm T} := {\bm T}^\dagger {\bm \O} = (\bar {\bm T}^A) 
= (\bar {\bm T}^{\hal}, - {\bar {\bm T}}^i )~, \qquad
\bar {\bm T}^i := ({\bm T}_i)^*~.
\eea
The superconformal group acts on ${\bm T}_A$ and $\bar {\bm T}^A$ as follows:
\bea
{\bm T}_A ~\to ~ g_A{}^B {\bm T}_B~, \qquad 
\bar {\bm T}^A~\to~ \bar {\bm T}^B (g^{-1})_B{}^A~.
\eea

\subsection{Supertwistor realization}
In complete analogy with the bosonic construction described in the previous section, 
compactified Minkowski superspace\footnote{The case $\cN=4$ is known to be somewhat special. 
Since off-shell supersymmetric theories exist for $\cN=1,2,3$, here we do not dwell 
on the special features of  $\cN=4$.}  
$\overline{\mathbb M}{}^{4|4\cN}$
is defined to be the space of null two-planes 
through the origin in  ${\mathbb C}^{4|\cN}$ \cite{Manin}.
Given such a two-plane, it is generated by two  supertwistors 
${\bm T}^\m  = ({\bm T}_A{\,}^\m )$, with $\m=1,2$, 
such that (i) the bodies\footnote{See \cite{BK} for the necessary information about
infinite dimensional Grassmann algebra $\L_\infty$.}
of ${\bm T}_{\hal}{\,}^1$ and  ${\bm T}_{\hal}{\,}^2$
are linearly independent twistors; and 
(ii) these supertwistors  obey the equations
\bea
\langle {\bm T}^\m, {\bm T}^\n \rangle \equiv \bar {\bm T}^\m {\bm T}^\n= 0~, \qquad 
\m ,\n =1,2~.
\label{3.6}
\eea 
The basis  chosen, $\{T^\m \}$, is defined  modulo 
the equivalence relation 
\bea
\{ {\bm T}^\m \}~ \sim ~ \{ \tilde{\bm T}^\m \} ~, \qquad
\tilde{\bm T}^\m = {\bm T}^\n\,C_\n{}^\m~, 
\qquad C \in {\sGL}(2,{\mathbb C}) ~.
\label{3.7}
\eea
The two bases, $\{{\bm T}^\m \}$ and $\{ \tilde{\bm T}^\m \}$, 
define one and the same two-plane in ${\mathbb C}^{4|\cN}$.
It may be shown that $\overline{\mathbb M}{}^{4|4\cN}$ is a homogeneous 
space of ${\sSU}(2,2|\cN) $. 

The standard $\cN$-extended Minkowski superspace, ${\mathbb M}^{4|4\cN} $ or 
more traditionally ${\mathbb R}^{4|4\cN}$,
can be identified with a certain open domain of $\overline{\mathbb M}{}^{4|4\cN}$ on which the upper 
$2\times 2$ matrix block of the supermatrix 
\bea
({\bm T}_{A}{\,}^\m)
= \left(
\begin{array}{c}
{\bm T}_\a{}^\m  \\
{\bm T}^{\dt \a \m}\\
{\bm T}_i{\,}^\m 
\end{array}
\right) 
\eea
is non-singular. The freedom to choose the basis, eq. \eqref{3.7}, can be used 
to fix $T_\a{}^\m = \d_\a{\,}^\m$. In this gauge, the supermatrix $({\bm T}_A{\,}^\m)$ takes the form 
have 
\bea
({\bm T}_A{\,}^\m)= \left(
\begin{array}{r}
\d_\a{}^\b \\  -{\rm i}\, {x}_+^{\dt \a \b}
 \\ 2 \,\q_i{}^\b
\end{array}
\right) 
~, \qquad  {x}_+^{\dt \a \b} :=  x_+^m \tilde{\s_m}^{\dt \a \b}~.
\label{super-two-plane-mod}
\eea
Due to \eqref{3.6}, the bosonic $\tilde{x}_+ =({x}_+^{\dt \a \b})$ 
and fermionic $\q = (\q_i{}^\b)$ variables 
obey the reality condition
\be
\tilde{x}_+ -\tilde{x}_- =4{\rm i}\, \q^\dagger \,\q~,
\qquad \tilde{x}_- = (\tilde{x}_+)^\dagger~.
\label{chiral}
\ee
It is solved by 
\be 
x_\pm^{\dt \a \b} = x^{\dt \a \b} \pm 2{\rm i} \, 
{\bar \q}^{\dt \a i} \q^\b_i ~,\qquad 
{\bar \q}^{\dt \a i} = ( \q^\a_i )^*~, 
\qquad \tilde{x}^\dagger = \tilde{x}~,
\ee
with $z = (x^a ,\q^\a_i , {\bar \q}_{\dt \a}^i)$ the coordinates
of $\cN$-extended Minkowski superspace ${\mathbb R}^{4|4\cN}$.
We  see that the supertwistors in the Minkowski chart
(\ref{super-two-plane-mod}) are parametrized 
by the {\it chiral } coordinates $x^a_+$ and $\q^\a_i$.
More details on the supertwistor construction for $\overline{\mathbb M}{}^{4|4\cN}$
can be found, e.g.,  in \cite{K-compact}.

We can elaborate on implications of the construction presented. 
Using the two supertwistors ${\bm T}^\m$, which describe a point of 
 $\overline{\mathbb M}{}^{4|4\cN}$, we define the following supermatrix 
 \bea
 {\bm Y}_{AB} := {\bm T}_A{\,}^\m {\bm T}_{B}{\,}^\n
\ve_{\m\n}~, \qquad \ve_{\m\n} =-\ve_{\n\m} ~, \qquad \ve_{12}=-1~.
\label{3.12}
\eea
It is graded antisymmetric, 
\bea
 {\bm Y}_{AB} = - (-1)^{\e_A \e_B}  {\bm Y}_{BA}~,
 \label{3.13}
 \eea
 and defined modulo the equivalence relation
\bea
{\bm Y}_{AB} ~\sim ~c \, {\bm Y}_{AB}~, \qquad c = \det (C_{\hal}{}^{\hbe} ) \in {\mathbb C}-\{0\}~.
\eea
The important property of this supermatrix is
\bea
{\bm Y}_{[AB} {\bm Y}_{C\}D} =0 \quad \longleftrightarrow \quad 
{\bm Y}_{[AB} {\bm Y}_{CD \}} =0~,
\eea 
where $[...\}$ denotes the graded antisymmetrization of indices. 

It should be emphasized that the supermatrix $ {\bm Y}_{AB} $ defined by \eqref{3.12} is non-zero, 
and so is the body of its bosonic block  ${\bm Y}_{\hal \hbe}$.
This follows from an easily verified statement: if ${\bm T}_A$ and ${\bm S}_A$ are two supertwistors 
such that the bodies of ${\bm T}_{\hal }$ and ${\bm S}_{\hal}$ are non-zero, 
then it holds that
\bea
{\bm T}_{[ A} {\bm S}_{B \} } = 0 \quad \longleftrightarrow \quad 
 {\bm T}_A = \l  {\bm S}_A~. 
\eea
If the bodies of ${\bm T}_{\hal }$ and ${\bm S}_{\hal}$ vanish, however, it is easy to construct two supertwistors ${\bm T}_A$ and ${\bm S}_A$ such that ${\bm T}_{[ A} {\bm S}_{B \} } =$ 0 but 
$ {\bm T}_A \neq  \l  {\bm S}_A$ for any $\l$.

In the Minkowski chart, choosing the gauge \eqref{super-two-plane-mod} gives
\bea
{\bm Y}_{AB}= \left(
\begin{array}{c|c||c} \phantom{\Big|}
\ve_{\a\b} & -{\rm i}\, {x}_{+\a}{\,}^{\dt  \b} & 2 \q_{\a j} \\
\hline \phantom{\Bigg|}
{\rm i}\, {x}_+{}^{\dt \a }{\,}_\b~ & ~\hf \ve^{\dt \a \dt \b}  {x}_+{}^{\dt \g \g} x_{+ \g \dt \g}~ & 
~ -2 {\rm i}\, {x}_+{}^{\dt \a \g} \q_{\g j} \\
\hline \hline \phantom{\Bigg|}
-2 \q_{i\b} & 2 {\rm i}\, \q_{i\g} {x}_+{}^{\dt \b \g} & 4 \q_i \q_j
\end{array}
\right) ~.
\label{chirall}
\eea

Using the dual supertwistors $\bar {\bm T}^\m$ allows us to define another supermatrix
\bea
\bar {\bm Y}^{AB} := \ve_{\m \n} \bar {\bm Y}^{\m \, A} \bar {\bm Y}^{\n \, B} ~.
\eea
The supermatrices $ {\bm Y}= ( {\bm Y}_{AB} )$ and $ \bar {\bm Y} =  (\bar {\bm Y}^{AB} )$
are related to each other as
\bea 
\bar  {\bm Y} = - {\bm \O}  {\bm Y}^\dagger {\bm \O}~.
\label{3.18}
\eea
where the definition of ${\bm Y}^\dagger$ is the same  as for ordinary matrices. 
The null conditions \eqref{3.6} give 
\bea
\bar {\bm Y}^{AC} {\bm Y}_{CB} =0~.
\eea

\subsection{Bi-supertwistor realization}\label{subsec3.2}

We give an alternative realization of compactified Minkowski superspace 
 $\overline{\mathbb M}{}^{4|4\cN}$. 
 In the space of all complex graded antisymmetric matrices, 
 $ {\bm Y}_{AB} = - (-1)^{\e_A \e_B}  {\bm Y}_{BA}$,
 we consider a surface $\frak L$ consisting of those supermatrices 
which obey the algebraic constraints \begin{subequations}
\bea
 {\bm Y}_{[AB} {\bm Y}_{CD \}} &=&0~, \label{3.20a}\\
\bar {\bm Y}^{AC} {\bm Y}_{CB} &=&0~, \label{3.20b}
\eea
\end{subequations}
and satisfy the following condition: for each supermatrix ${\bm Y}\in \frak L$,
 the {\it body} of its bosonic block 
${\bm Y}_{\hal \hbe}$ defined by 
\bea
{\bm Y} =
\left(
\begin{array}{cc}
{\bm Y}_{\hal \hbe} & {\bm Y}_{\hal j }  \\
{\bm Y}_{i \hbe}  & {\bm Y}_{ij} 
\end{array}
\right) ~
\eea
is a non-zero antisymmetric $4\times 4$ matrix. 
Our goal is to show that the superspace $\overline{\mathbb M}{}^{4|4\cN}$
can be identified with the  space of equivalence classes in ${\frak L}$ with respect to 
the equivalence relation 
\bea
{\bm Y}_{AB} ~\sim ~c \, {\bm Y}_{AB}~, \qquad c  \in {\mathbb C}-\{0\}~.
\label{3.22}
\eea
It is natural to think of ${\bm Y}_{AB}$
as a graded two-form on the dual supertwistor space. 

Let us first demonstrate that eq. \eqref{3.20a} implies that $ {\bm Y}_{AB} $ is decomposable provided 
the body of ${\bm Y}_{\hal \hbe}$ is  non-zero. This means 
that  $ {\bm Y}_{AB} $ can be represented as
\bea
{\bm Y}_{AB} = {\bm T}_{A}{\bm S}_{B} - (-1)^{\e_A \e_B} {\bm T}_{B}{\bm S}_{A}~,
\label{3.23}
\eea
for some supertwistors ${\bm T}_{A}$ and ${\bm S}_{A}$. 
It follows from  eq. \eqref{3.20a}  that 
\bea
 {\bm T}_{[A} {\bm Y}_{BC \}} &=&0~, 
 \label{3.24}
 \eea
 where ${\bm T}_A:= {\bm V}^B{\bm Y}_{BA}$,
 for any dual supertwistor ${\bm V}^A$. Since the body of ${\bm Y}_{\hal \hbe}$ is  non-zero, 
 it is possible to choose ${\bm V}^A$ such that the body of ${\bm T}_{\hal}$ is non-zero. 
The properties of ${\bm T}_{A} $ and ${\bm Y}_{AB} $ are such that we can apply a generalization of Cartan's lemma.\footnote{For completeness, we recall the formulation of Cartan's  lemma, see e.g.
 \cite{WW,Sternberg}. Consider a finite 
dimensional vector space $V$. Let $\o \in \wedge^pV $ be a $p$-vector, and $\vf \in V$ be a non-zero one-vector 
such that $\vf \wedge \o =0$. Then there exists a $(p-1)$-vector $\eta$ such that $\o = \vf \wedge \eta$.} 
This generalization states that the condition \eqref{3.24} implies the validity of \eqref{3.23}, for 
some supertwistor ${\bm S}_{A} $ such that the body of ${\bm S}_{\hal}$ is non-zero. 
It is clear that the bodies of ${\bm T}_{\hal}$ and ${\bm S}_{\hal}$ are linearly independent twistors.
Since ${\bm Y}_{AB} $ is defined modulo arbitrary re-scalings,  eq. \eqref{3.22}, 
one can see that  the two supertwistors ${\bm T}^\m := ({\bm T}, {\bm S})$ are defined modulo 
the equivalence relation \eqref{3.7}.

Starting from the graded antisymmetric supermatrix \eqref{3.23}, we introduce its conjugate
$\bar {\bm Y}^{AB}$, eq. \eqref{3.18}, and compute the left-hand side of \eqref{3.20b}. 
Then eq. \eqref{3.20b} becomes equivalent to 
\bea
\bar {\bm Y}^{AC} {\bm Y}_{CB} =
 \bar {\bm T}^{A} \Big\{ 
\langle {\bm S}, {\bm T} \rangle {\bm S}_B &-& \langle {\bm S}, {\bm S} \rangle {\bm T}_{B} \Big\} \non \\
&+&  \bar {\bm S}^{A} \Big\{ 
\langle {\bm T}, {\bm S} \rangle {\bm T}_{B} - \langle {\bm T}, {\bm T} \rangle {\bm S}_{B} \Big\} =0~.
\eea
Since the bodies of $ \bar {\bm T}^{\hal}$ and $ \bar {\bm S}^{\hal}$
are linearly independent, we conclude that the two expressions in figure brackets must vanish. 
Since the bodies of $ {\bm T}_{\hal}$ and $  {\bm S}_{\hal}$ 
are linearly independent, we end up with the null conditions
\bea
 \langle {\bm T}, {\bm T} \rangle =  \langle {\bm T}, {\bm S} \rangle =  \langle {\bm S}, {\bm T} \rangle
=  \langle {\bm S}, {\bm S} \rangle = 0~.
\eea
As a result, we have demonstrated that the bi-supertwistor realization introduced
 is completely equivalent to the standard supertwistor realization of 
$\overline{\mathbb M}{}^{4|4\cN}$ described in the previous subsection. 
  
The constraints  \eqref{3.20a} and \eqref{3.20b} for $\cN=1$ were identified in the third version of \cite{GSS}. 
Ref. \cite{Maio} closely followed the first version of \cite{GSS} and did not provide the correct constraints.
The constraint \eqref{3.20a} and a considerable part of the bi-supertwistor 
construction for general $\cN$, including eq. \eqref{chirall}, were known to Siegel 
in the mid 1990s \cite{Siegel93,Siegel95}. 
The equivalence between the supertwistor and bi-supertwistor formulations was not proved in 
\cite{GSS,Siegel93,Siegel95}.

\section{Compactified 4D $\cN=2$ harmonic/projective superspace}
 
As is well known, all  $\cN=2$ supersymmetric  theories in four dimensions 
are naturally formulated in the superspace ${\mathbb R}^{4|8} \times {\mathbb C}P^1$ 
introduced for the first time by Rosly \cite{Rosly}. 
Depending on the specific superfield methods employed to construct off-shell $\cN=2$ supersymmetric theories, 
this superspace  is called harmonic \cite{GIKOS,GIOS} or projective \cite{KLR,LR}. 
Here we start by describing the conformally compactified version of  ${\mathbb R}^{4|8} \times {\mathbb C}P^1$ 
following Ref. \cite{K-compact} which built on the earlier publications  \cite{Rosly2,LN,HH}.
After that we introduce a new bi-supertwistor realization for this compactified superspace. 
 
 Ordinary supertwistors introduced by Ferber \cite{Ferber}, 
 \bea 
({\bm T}_A) = \left(
\begin{array}{c}
{\bm T}_{\hal}  \\
{\bm T}_i 
\end{array}
\right) ~,
\eea
are characterized by  the following Grassmann parities of their components: 
\bea
  \e( {\bm T}_A ) = \e_A = \left\{  
\begin{array}{c}
 0 \qquad A=\hal  \\
 1 \qquad A=i
\end{array}
\right.~.
\eea
Such supertwistors are often called {\it even}.
One can also consider {\it odd} supertwistors, 
\bea 
( {\bm \X}_A) = \left(
\begin{array}{c}
{\bm \X}_{\hal}  \\
{\bm \X}_i 
\end{array}
\right) ~,
 \eea
 with opposite Grassmann parities 
 \bea
  \e ( {\bm \X}_A) 
= 1 +\e_A   \qquad ( {\rm mod} ~2) ~.
\eea
Both even and odd supertwistors should be used \cite{Rosly2,LN} in order 
to define harmonic-like superspaces in extended conformal supersymmetry.

Now, we accompany the two even supertwistors ${\bm T}^\m$, 
which occur in the construction of  the compactified
$\cN=2 $ superspace $\overline{\mathbb M}{}^{4|8} $,
by an odd supertwistor $\bm \X$ such that  the body of ${\bm \X}_i$ is non-zero. 
These supertwistors are required to obey the null conditions 
\be
\langle {\bm T}^\m, {\bm T}^\n \rangle = \langle {\bm T}^\m, {\bm \X} \rangle = 
0~, \qquad 
\m, \n =1,2 ~,
\label{nullplane3}
\ee 
and are defined modulo the equivalence  relation
\bea
({\bm \X}, {\bm T}^\m)~\sim ~  ({\bm \X}, {\bm T}^\n) \,
\left(
\begin{array}{cc}
 d~  &0  \\  
 \r_\n~ & C_\n{}^\m 
\end{array}
\right) ~,\qquad 
\left(
\begin{array}{cc}
 d~  &0  \\  
 \r~ & \cC
\end{array}
\right) \in {\sGL}(1|2)~,
\label{4.6}
\eea
with $\r_\n$  anticommuting complex parameters.
The superspace obtained can be seen to be 
$\overline{\mathbb M}{}^{4|8} \times {\mathbb C}P^1$. It is a homogeneous space 
of ${\sSU}(2,2|2) $.

To understand the global structure of the superspace introduced, 
it is convenient to restrict our consideration to its 
Minkowski chart defined the same way as in the previous section. 
The freedom to perform equivalence transformations \eqref{4.6}
allows us to choose ${\bm T}^\m$ and ${\bm \X}$ to look like
\bea
({\bm T}_A{\,}^\m)= \left(
\begin{array}{r}
\d_\a{}^\b \\  -{\rm i}\, {x}_+^{\dt \a \b}
 \\ 2 \,\q_i{}^\b
\end{array}
\right) ~, 
\qquad 
 ({\bm \X}_A)  
=  \left(
\begin{array}{c}
 0 \\   2{\bar \q}^{\dt \a j} v_j
 \\  v_i
\end{array}
\right) ~, \qquad v_i \in {\mathbb C}^2 -\{ 0\} ~.
\eea
  The isotwistor $v_i$ is defined modulo the equivalence relation
\bea
v_i ~\sim ~d\, v_i~, \qquad d \in {\mathbb C}-\{ 0\} ~.
\eea
This shows that the  superspace under consideration is indeed 
$\overline{\mathbb M}{}^{4|8} \times {\mathbb C}P^1$.

We are in a position to formulate a bi-supertwistor realization for 
$\overline{\mathbb M}{}^{4|8} \times {\mathbb C}P^1$. It is given in terms of 
complex variables 
$(  {\bm Y}_{AB} , {\bm \X}_A)$, where ${\bm Y}_{AB}$
obeys the conditions given in subsection \ref{subsec3.2}, 
while  $\bm \X$ is an odd supertwistor such that  
(i) the body of ${\bm \X}_i$ is non-zero;\footnote{This condition implies that
the body of $\bar {\bm \X}^C  {\bm \X}_C$ is non-zero, that is
$(\bar {\bm \X}^C  {\bm \X}_C )^{-1}$ is well defined.} and (ii) $\bm \X$ obeys the condition
 \bea
 \bar {\bm Y}^{AB} {\bm \X}_B =0~.
\eea 
The variables $(  {\bm Y}_{AB} , {\bm \X}_A)$ are defined modulo 
the equivalence relation 
\bea
( {\bm \X}_A ,  {\bm Y}_{AB} ) ~\sim ~ ( {\bm \X}'_A ,  {\bm Y}'_{AB} )= 
(  {\bm \X}_A , {\bm Y}_{AC} )
\left(
\begin{array}{cc}
 d ~  &0  \\  
\r^C~ & c \,\d^C{}_B
\end{array}
\right) ~,
\label{4.10}
\eea
where $ c, d \in {\mathbb C}-\{ 0\} $, and
$\r^C$ is an odd dual supertwistor.  

Let us introduce a superfield of the general form ($n\geq 0$):
\bea
\F^{(n)} ({\bm Y}, \bar {\bm Y} , {\bm \X}, \bar {\bm \X} ) = \sum_{k=0}^{\infty}
\frac{\bar {\bm \X}^{B_k} \dots \bar {\bm \X}^{B_1} } {(    \bar {\bm \X}^C  {\bm \X}_C)^k}  \,
\F_{B_1 \dots B_k} {}^{A_1 \dots A_{n+k} } ({\bm Y}, \bar {\bm Y} ) \,
{\bm \X}_{A_{n+k}} \dots {\bm \X}_{A_1}~,~~~
\label{4.11}
\eea
where the Fourier coefficients $\F_{B_1 \dots B_k} {}^{A_1 \dots A_{n+k} } ({\bm Y}, \bar {\bm Y} )$ 
have the following properties: (i) they
are homogeneous functions of $\bm Y$'s and their conjugates, 
\bea
\F_{B_1 \dots B_k} {}^{A_1 \dots A_{n+k} } ( c {\bm Y},  \bar c \bar {\bm Y} )
= c^{-\D} \, {\bar c}^{{}\,-{\bar \D} } \,\F_{B_1 \dots B_k} {}^{A_1 \dots A_{n+k} } ({\bm Y}, \bar {\bm Y} )~,
\eea
for some parameter $\D$ and $\bar \D$;\\
(ii) they 
are graded antisymmetric in their $A$-indices and separately in their $B$-indices, 
\bea
\F_{B_1 \dots B_k} {}^{A_1 \dots A_{n+k} } ({\bm Y}, \bar {\bm Y} ) 
= \F_{ [ B_1 \dots B_k \}} {}^{[A_1 \dots A_{n+k} \}} ({\bm Y}, \bar {\bm Y} ) ~;
\eea
(iii) they are tensors at the point $( {\bm Y} , \bar {\bm Y })$ of  $\overline{\mathbb M}{}^{4|8} $, that is 
\begin{subequations}
\bea
\bar {\bm Y}^{CD}
\F_{D B_2\dots B_k} {}^{A_1 \dots A_{n+k} } ({\bm Y}, \bar {\bm Y} ) &=&0~, \\
\F_{B_1 \dots B_k} {}^{A_1 \dots A_{n+k-1} D } ({\bm Y}, \bar {\bm Y} ) {\bm Y}_{DC}&=&0~.
\eea
\end{subequations}
These tensor conditions guarantee that $\F^{(n)} ({\bm Y}, \bar {\bm Y} , {\bm \X}, \bar {\bm \X} ) $
changes under the equivalence transformation \eqref{4.10} as follows:
\bea
\F^{(n)} ({\bm Y}', \bar {\bm Y}' , {\bm \X}', \bar {\bm \X}' ) = d^n\,
 c^{-\D} \, {\bar c}^{{}\,-{\bar \D} } \,
 \F^{(n)} ({\bm Y} , \bar {\bm Y} , {\bm \X} , \bar {\bm \X} )  ~.
\eea
Eq. \eqref{4.11} defines a superconformal  Fourier expansion. 
Without loss of generality, the Fourier coefficients in  \eqref{4.11} can be subject to an irreducibility condition 
that a super-trace of any $A$-index with a $B$-index vanish. 

It should be mentioned that there exist two more equivalent realizations for  the superspace
$\overline{\mathbb M}{}^{4|8} \times {\mathbb C}P^1$, which will be referred to 
as Type A and Type B formulations.
Both realizations are given in terms of complex variables 
$(  {\bm Y}_{AB} , {\bm \X}_A, {\bm \S}_A)$, where ${\bm Y}_{AB}$
obeys the conditions given in subsection \ref{subsec3.2}, 
while  $\bm \X$  and $\bm \S$ are odd supertwistors such that  
(i) the bodies of ${\bm \X}_i$  and ${\bm \S}_i$ are non-zero;
and (ii) $\bm \X$ and $\bm \S$ obey the null conditions
\bea
\bar {\bm Y}^{AB} {\bm \X}_B =0~, \qquad  \bar {\bm Y}^{AB} {\bm \S}_B =0~; 
\eea 
(iii) the odd supertwistor $\bm \X$ is defined modulo arbitrary equivalence 
transformations \eqref{4.10}. The two realizations differ in additional conditions (iv) and (v)
imposed 
on $\bm \S$. Let us describe these conditions. 

In Type A formulation,  the odd supertwistor $\bm \S$ is required to (iv) obey 
the additional null condition 
\bea
\bar {\bm \S}^{B} {\bm \X}_B =0~;
\eea
as well as (v) be defined modulo the equivalence relation
\bea
 {\bm \S}_A  ~\sim ~  {\bm \S}'_A = f  {\bm \S}_A + {\bm Y}_{AC} \k^C~, 
 \label{B.3}
\eea
where $ f \in {\mathbb C}-\{ 0\} $, and
$\k^C$ is an arbitrary odd dual supertwistor.  
It follows that the bodies of  ${\bm \X}_i$  and ${\bm \S}_i$ are linearly independent.
One can see that no additional degrees of freedom are associated with $\bm \S$.

In Type B formulation,  the odd supertwistor $\bm \S$ is such that (iv) the bodies of 
${\bm \X}_i$  and ${\bm \S}_i$ are linearly independent; and (v) $\bm \S$ is defined modulo 
the equivalence relation\footnote{If the equivalence relation \eqref{B.4} is replaced by
\eqref{B.3}, then we end up with  a space  $\overline{\mathbb M}{}^{4|8} \times T^*{\mathbb C}P^1$, 
where $T^*{\mathbb C}P^1$ denotes the cotangent bundle of ${\mathbb C}P^1$. One can think of 
$T^*{\mathbb C}P^1$ as the complexification of ${\mathbb C}P^1$.  The importance of 
superspace  ${\mathbb R}{}^{4|8} \times T^*{\mathbb C}P^1$ in the context $\cN=2$ supersymmetric 
sigma models has recently been emphasized by Butter \cite{Butter}.} 
\bea
{\bm \S}_A  ~\sim ~  {\bm \S}'_A = f  {\bm \S}_A + g \,{\bm \X}_A
+{\bm Y}_{AC} \k^C~, 
\label{B.4}
\eea
where $ f \in {\mathbb C}-\{ 0\} $, $ g \in {\mathbb C} $,  and
$\k^C$ is an arbitrary odd dual supertwistor.  
As in the previous case of Type A formulation, no degrees of freedom are associated with $\bm \S$.
Type B formulation is the bi-supertwistor version of the so-called harmonic realization for 
$\overline{\mathbb M}{}^{4|8} \times {\mathbb C}P^1$ introduced in \cite{K-compact}.

\section{Compactified 3D Minkowski space} 

In the remainder of this paper, we present three-dimensional analogues of the four-dimensional results discussed in sections 2 to 4. The (super)twistor realizations for 3D compactified Minkowski space and its supersymmetric extensions were developed in \cite{KPT-MvU}. 
Here we will build on the constructions presented in \cite{KPT-MvU}.
The interested reader is referred to that paper for more details, including the spinor conventions in $3+2$ dimensions. 

\subsection{Twistor realization} 
Consider a symplectic four-dimensional real vector space. We can think of it as  ${\mathbb R}^4$ 
equipped with a skew-symmetric inner product:
\be
\langle T | S \rangle_{J}: = T^{\rm T}J \, S  \equiv T_{\hat \a} J^{\hat \a \hat \b} S_{\hat \b}
=- \langle S| T \rangle_{J}
~, \qquad
J =\big(J^{\hat \a \hat \b} \big) 
=\left(
\begin{array}{cc}
0  & {\mathbbm 1}_2\\
 -{\mathbbm 1}_2  &    0
\end{array}
\right) ~,
\label{J}
\ee
for any  vectors 
$T,S\in {\mathbb R}^4$. By construction, this inner product is invariant under the group\footnote{This 
group was denoted   $\sSp(2,{\mathbb R})$ in \cite{KPT-MvU}.}
  $\sSp(4,{\mathbb R})$.  We refer to this vector space as the 3D twistor space.
Its elements $T,S\in {\mathbb R}^4$ are  called 3D twistors.
 
The elements of the group
$\sSp(4,{\mathbb R})$
 can  be represented by $4\times 4$ block 
matrices
\bea
g=\big(g_{\hat \a}{}^{\hat \b}\big)= \left(
\begin{array}{cc}
 \cA  &- \cB\\
-\cC &    \cD 
\end{array}
\right) \in \sSL(4,{\mathbb R}) ~, \qquad 
g^{\rm T} \,J \,g = J~,
\label{SP(2)}
\eea
where  $\cA,\cB,\cC$ and $\cD$ are $2\times 2$ matrices. The symplectic group $\sSp(4,{\mathbb R})$ is the 
two to one  covering of the connected component,  $\sSO_0 (3,2)$, of the
conformal group in three dimensions. 
A twistor looks like 
\bea
T = (T_{\hal}) =\left(
\begin{array}{c}
 f_\a \\
  h^\a 
  \end{array}
\right)~,\eea
with the two-component vectors $f_\a$ and $h^\a$ being real commuting 3D  spinors. 
   
A {\it Lagrangian subspace} of the twistor space
is defined to be a maximal isotropic vector subspace of ${\mathbb R}^4$.
Such a subspace is necessarily two-dimensional. We denote by $\overline{\mathbb M}{}^3_{\rm T}$ 
the space of all 
Lagrangian subspaces of ${\mathbb R}^4$. One can show that  $\overline{\mathbb M}{}^3_{\rm T}$
is a homogeneous space  of the group $\sSp(4,{\mathbb R})$ and has the structure 
\bea
\overline{\mathbb M}{}^3_{\rm T} =\sU (2) / \sf{O} (2)~, 
\eea
 see, e.g., \cite{KPT-MvU,Berndt} for technical details. 

Conformally compactified 3D Minkowski space can be identified  with $\overline{\mathbb M}{}^3_{\rm T}$.
Indeed, given a Lagrangian subspace, 
it is generated by two linearly independent twistors $T^\m$, with $\m=1,2$,
such that
\be
\langle T^1| T^2 \rangle_J = 0~. 
\label{nullplane1}
\ee 
Obviously, the basis  chosen, $\{T^\m\}$, is defined only modulo 
the equivalence relation 
\be
\{ T^\m \} ~ \sim ~ \{ \tilde{T}^\m \} ~, \qquad
\tilde{T}^\m = T^\n\,R_\n{}^\m~, 
\qquad R \in \sGL(2,{\mathbb R}) ~.
\label{nullplane2}
\ee
Minkowski space ${\mathbb M}{}^3 \equiv {\mathbb R}^{2,1}$
can be identified with an open dense subset of  $\overline{\mathbb M}{}^3_{\rm T}$ consisting of
 those Lagrangian subspaces which are described by $4\times 2$ matrices of the form:
\bea
(T_{\hal}{\,}^\m ) 
=  \left(
\begin{array}{c}
 F_\a{\,}^\m\\  H^{\a \, \m} 
\end{array}
\right) 
~, \qquad \det F \neq 0~.
\eea
In accordance with the equivalence relation (\ref{nullplane2}), we can choose $F = {\mathbbm 1}_2$. 
Then the null condition gives
\bea
(T_{\hal}{\,}^\m ) = 
\left(
\begin{array}{r}
 {\mathbbm 1}_2\\  -x
\end{array}
\right)  ~, \qquad 
x^{\rm T} =x = (x^{\a \b})  \in \sMat (2, {\mathbb R} )~.
\eea
This is a standard matrix realization of 3D Minkowski space.  
The conformal transformation \eqref{SP(2)} acts on $x$ as follows:
\bea
x' = (C+Dx)(A+Bx)^{-1}~.
\eea
The Poincar\'e group corresponds to the subgroup of  $\sSp(4,{\mathbb R})$ 
consisting of the matrices
\bea
g= \left(
\begin{array}{c | c }
  M  ~& ~ 0  \\
\hline 
-aM  ~& ~  (M^{-1})^{\rm T}
\end{array}
\right) ~, \qquad a= a^{\rm T}     \in \sMat (2, {\mathbb R} )~, \quad
M \in \sSL(2, {\mathbb R})~.
\eea

\subsection{Bi-twistor realization} 
  
We now describe an alternative, {\it bi-twistor} realization of  $\overline{\mathbb M}{}^3$. 
Let us denote by $\cL$ 
the set of all {\it non-zero} real antisymmetric matrices $X_{\hal \hbe} = - X_{\hbe \hal}$ 
which obey the algebraic constraints 
\begin{subequations} 
\bea
X_{[\hal \hbe} X_{\hga  \hde ]} &=&0~, 
\\ 
X_{\hal \hga} J^{\hga \hde} X_{\hde \hbe}&=&0~, \\
J^{\hbe \hal} X_{\hal \hbe}  &=&0.
\eea
\end{subequations}
We introduce the quotient space $\overline{\mathbb M}{}^3_{\text{BT}} = \cL / \! \!\sim$, 
where the equivalence relation is defined by 
\bea
X_{\hal \hbe} ~\sim ~r \, X_{\hal \hbe}~, \qquad r  \in {\mathbb R}-\{0\}~.
\eea

The twistor and bi-twistor realizations are related to each other as follows:
\bea
X_{\hal \hbe} := T_{\hal}{}^\m T_{\hbe}{}^\n \ve_{\m \n}~, \qquad \ve_{\m \n }= -\ve_{\n \m}~, \qquad
\ve_{12} =-1~.
\eea
The proof is left to the reader  as an exercise. 

\section{Compactified 3D Minkowski superspace} 

Consider the graded  symplectic metric on ${\mathbb R}^{4|\cN}$ 
\bea
{\bm J} =({\bm J}^{AB}) =  \left(
\begin{array}{cc}
J^{\hal \hbe} ~&~ 0 \\
0 ~& ~{\rm i} \,
\d^{IJ}
\end{array} \right) ~, \qquad I,J=1, \dots, \cN~.
\label{supermetric}
\eea
The 3D $\cN$-extended superconformal group  $\sOSp(\cN|4, {\mathbb R})$
consists of supermatrices $g$ of the form
\bea
g^{\rm sT} {\bm J}\, g = {\bm J} ~, \qquad
g= \left(
\begin{array}{c||c}
 A  & B\\
 \hline \hline
C &    D 
\end{array}
\right) ~,
\qquad 
g^{\rm sT}=\left(
\begin{array}{c||r}
 A^{\rm T}  & -C^{\rm T}\\
\hline \hline
B^{\rm T}  &    D^{\rm T} 
\end{array}
\right) ~.
\label{6.2}
\eea
Here the even matrices $A, D$ and the odd matrix $B$  have real matrix elements, 
while the odd matrix $C$ has purely imaginary matrix elements. 
Supermatrices $g$ of this type are called {\it real}, in accordance with \cite{BK}.

The superconformal group $\sOSp(\cN|4, {\mathbb R})$ naturally acts on supertwistor space 
${\mathbb R}^{4|\cN}$ spanned by elements of the form
\bea
{\bm T} \equiv ({\bm T}_A) = 
\left(
\begin{array}{c}
 {\bm T}_{\hal} \\
  {\rm i} \, \vf_I
  \end{array}
\right)
= \left(
\begin{array}{c}
 f_\a \\
  h^\a \\
  {\rm i} \, \vf_I
  \end{array}
\right)~, \qquad 
 \e( {\bm T}_A ) = \e_A = \left\{  
\begin{array}{c}
 0 \qquad A=\hal  \\
 1 \qquad A=I
\end{array}
\right.
\label{6.3}
\eea
 and endowed with the graded symplectic two-form 
 ${\bm \cJ}  = \hf {\bm J}^{AB} \,\rd {\bm T}_B \wedge \rd {\bm T}_A$.
 This action preserves  ${\bm \cJ} $, and thus the symplectic inner product 
 on ${\mathbb R}^{4|\cN}$ defined by 
\bea
\langle {\bm T}| {\bm S} \rangle_{\bm J}: = {\bm T}^{\rm sT}{\bm J} \, {\bm S}
=- \langle {\bm S}| {\bm T} \rangle_{\bm J}
~,  \qquad
{\bm T}^{\rm sT} = \big( {\bm T}_{\hal}
\,, -   ~{\rm i} \,  \vf_I \big)~, 
\label{6.4}
\eea
with the graded symplectic matrix $\bm J$ defined in (\ref{supermetric}).
Any element ${\bm T} \in {\mathbb R}^{4|\cN}$ 
is called an {\it even real supertwistor}. 

\subsection{Supertwistor realization} 

A {\it Lagrangian subspace} of ${\mathbb R}^{4|\cN}$ is defined to be a maximal isotropic subspace
of the supertwistor space
\cite{KPT-MvU}. 
We denote by $\overline{\mathbb M}{}^{3|2\cN}$ the space of all 
Lagrangian subspaces of ${\mathbb R}^{4|\cN}$. 
Given such a subspace, it is generated by two  supertwistors 
${\bm T}^\m$ such that \\
${}\quad$(i) the bodies of ${\bm T}_{\hal}{\,}^1$ and $ {\bm T}_{\hal}{\,}^2$ are linearly independent twistors;\\
${}\quad$(ii)  ${\bm T}^1$ and $ {\bm T}^2$ obey the null condition
\be
\langle {\bm T}^1 | {\bm T}^2  \rangle_{\bm J}=0~;
\label{5.3}
\ee
${}\quad$(iii) ${\bm T}^1$ and $ {\bm T}^2$ are defined only modulo 
the equivalence relation 
\bea
\{ {\bm T}^\m \} ~ \sim ~ \{ \tilde{\bm T}^\m \} ~, \qquad
\tilde{\bm T}^\m = {\bm T}^\n\,R_\n{}^\m~, 
\qquad R \in \sGL(2,{\mathbb R}) ~.
\label{super-nullplane2}
\eea

A dense open subset ${\mathbb M}{}^{3|2\cN}$ of  $\overline{\mathbb M}{}^{3|2\cN}$ consists of
those Lagrangian subspaces which are described by supermatrices of the form
\bea
\big( {\bm T}_A{\,}^\m\big)=
\left(
\begin{array}{c}
 F_\a{\,}^\m \\  H^{\a\, \m} \\ {\rm i}\, \U_I{\,}^\m
\end{array}
\right)  ~, \qquad \det ( F_\a{\,}^\m) \neq 0~.
\eea
Using the equivalence relation (\ref{super-nullplane2}) allows us to choose $F={\mathbbm 1}_2$, and hence
\bea
\big( {\bm T}_A{\,}^\m\big)=
\left(
\begin{array}{c}
 \d_\a{}^\b \\  -x^{\a\b} +\frac{\rm i}{2}\ve^{\a\b} \q^2  \\ {\rm i}\sqrt{2}\,\q_I{}^\b
\end{array}
\right) ~, \qquad x^{\a\b} =x^{\b\a}~,
\quad \q^2:= \q^\a_I \q_{\a I} ~.~~~
\label{5.8}
\eea
Here the bosonic $x^{\a\b}$ and fermionic $\q_I^\a \equiv \q_I{}^\a$ parameters are real.
Therefore, the subset  ${\mathbb M}{}^{3|2\cN} \subset \overline{\mathbb M}{}^{3|2\cN}$ introduced
can be identified with  ${\mathbb R}^{3|2\cN}$.
This subset is a transformation space of the 3D
$\cN$-extended super-Poincar\'e group, 
$ {\mathfrak P}(3|\cN)$, which is  a subgroup of the superconformal group 
$\sOSp(\cN|2, {\mathbb R})$, eq. \eqref{6.2}, 
spanned by group elements
of  the form:
\begin{subequations} 
\bea
g &=& s(a, \e) \, h( M) ~, \\
s(a, \e) &=& 
\left(
\begin{array}{c | c ||c}
  \d_\a{}^\b  ~& ~ 0~ &~0   \\
\hline 
-{a}^{\a \b} +\frac{\rm i}{2}\ve^{\a\b} \e^2 ~& ~  \d^\a{}_\b~&~ -\sqrt{2}\e^\a{}_J
\\
\hline
\hline
{\rm i}\sqrt{2}\, \e_I{}^\b ~& ~0~&~\d_{IJ}
\end{array}
\right) 
~, 
\label{B.1b}
\\
h(M) &=&  \left(
\begin{array}{c | c ||c}
  M  ~& ~ 0~ &~0   \\
\hline 
0  ~& ~  (M^{-1})^{\rm T}~&~ 0
\\
\hline
\hline
0 ~& ~0~&~{\mathbbm 1}_\cN
\end{array}
\right) ~, \qquad
M \in \sSL(2, {\mathbb R})~.
\label{B.1c}
\eea
\end{subequations}
In eq. (\ref{B.1b}), the bosonic  ($a^{\a \b}=a^{\b \a} $) and fermionic 
($\e_I{}^\a = \e^\a{}_I \equiv \e^\a_I$) parameters are real.
Evaluating the action of $ {\mathfrak P}(3|\cN)$ on ${\mathbb M}{}^{3|2\cN}$ shows that this space is 
3D $\cN$-extended Minkowski superspace.

\subsection{Bi-supertwistor realization} \label{subsection6.2}
  
We give an alternative, {\it bi-supertwistor} realization of compactified Minkowski superspace 
 $\overline{\mathbb M}{}^{3|2\cN}$. 
  In the space of all  {\it real} graded antisymmetric matrices, 
 $ {\bm X}_{AB} = - (-1)^{\e_A \e_B}  {\bm X}_{BA}$,
 we consider a surface $\frak L$ consisting of those supermatrices 
which obey the algebraic constraints \begin{subequations}
\bea
 {\bm X}_{[AB} {\bm X}_{CD \}} &=&0~, 
 \\
(-1)^{\e_C} {\bm X}_{AC} {\bm J}^{CD} {\bm X}_{DB} &=&0~, \\
{\bm J}^{BA} {\bm X}_{AB} &=&0,
\eea
\end{subequations}
and satisfy the additional condition: for each supermatrix ${\bm X}\in \frak L$,
 the {\it body} of its bosonic block 
${\bm X}_{\hal \hbe}$ defined by 
\bea
{\bm X} =
\left(
\begin{array}{cc}
{\bm X}_{\hal \hbe} & {\bm X}_{\hal J }  \\
{\bm X}_{I \hbe}  & {\bm X}_{IJ} 
\end{array}
\right) ~
\eea
is a non-zero antisymmetric $4\times 4$ matrix. 
The supermatrix $\bm X$ must be real in the following sense \cite{BK}
\bea
({\bm X}_{AB})^* = (-1)^{\e_A + \e_B + \e_A \e_B} {\bm X}_{AB}~.
\eea
It turns out that compactified Minkowski superspace
can be identified with the quotient space $\overline{\mathbb M}{}^{3|2\cN}_{\rm BT} = {\frak L} /\sim$, 
where the equivalence relation is defined by 
\bea
{\bm X}_{AB} ~\sim ~r \, {\bm X}_{AB}~, \qquad r  \in {\mathbb R}-\{0\}~.
\eea
  
The supertwistor and bi-supertwistor realizations are related to each other as follows:
 \bea
 {\bm X}_{AB} := {\bm T}_A{\,}^\m {\bm T}_{B}{\,}^\n
\ve_{\m\n}~, \qquad \ve_{\m\n} =-\ve_{\n\m} ~, \qquad \ve_{12}=-1~.
\eea
This can be proved in complete analogy with the four-dimensional case considered 
in section 3. 

In the Minkowski chart, choosing the gauge \eqref{5.8} gives
\begin{align}
{\bm X}_{AB}= \left(
\begin{array}{c|c||c} \phantom{\Big|}
\ve_{\a\b} & - {x}_{\a}{}^{ \b} -\frac{\ri}{2} \d_\a{}^\b \q^2 & \ri \sqrt{2}\q_{\a J} \\
\hline 
\phantom{\Bigg|}
{x}{}^{\a }{}_\b  +\frac{\ri}{2} \d^\a{}_\b  \q^2~ & ~ \hf \ve^{ \a  \b} \Big(  {x}{}^{\g \d} x_{ \g \d} +\hf (\q^2)^2\Big)~ & 
-  {\rm i} \sqrt{2} {x}{}^{ \a \g} \q_{\g J}  - \frac{1}{\sqrt{2}} \q^\a{}_J \q^2\\
\hline \hline 
-\ri \sqrt{2}\q_{I\b}  
  \phantom{\Bigg|}   
& {\rm i} \sqrt{2} \q_{I \g }{x}{}^{ \g \b}  + \frac{1}{\sqrt{2}} \q_I{}^\b \q^2  & -2 \q_I \q_J
\end{array}
\right) ~.
\end{align}

\section{Compactified $\cN$-extended harmonic/projective superspaces in three dimensions}

In \cite{Howe:1994ms,KPT-MvU},  new homogeneous spaces
of the superconformal group $\sOSp(\cN|2, {\mathbb R})$ were constructed 
that include  $\overline{\mathbb M}{}^{3|2\cN}$ as a submanifold. 
Their general structure is
$\overline{\mathbb M}{}^{3|2\cN} \times {\mathbb X}^\cN_m$, 
where $   {\mathbb X}^\cN_m$ is a compact manifold, 
for any integer $0<m\leq [\cN/2]$\footnote{As usual, the notation $[\cN/2]$ is used for
the integer part of $\cN/2$.}. Such superspaces are nontrivial and have interesting applications  
for $\cN >2$. The supertwistor formulation for $\overline{\mathbb M}{}^{3|2\cN} \times {\mathbb X}^\cN_m$
given in \cite{KPT-MvU} makes use of both even and odd supertwistors. 
An {\it odd} supertwistor looks like 
\bea
{\bm \S} = ({\bm \S}_A)= \left(
\begin{array}{c}
{\bm \S}_{\hal} \\
 {\bm \S}_I
\end{array}
\right)~, 
\qquad 
\e ( {\bm \S}_A ) =1 + \e_A \qquad ({\rm mod} ~2)
~.
\eea
This supertwistor is called real if all the components ${\bm \S}_A$ are real.
The super-transpose of $\bm \S$ is defined to coincide with the ordinary transpose,
\bea
{\bm \S}^{\rm sT} = \big( {\bm \S}_{\hal} , {\bm \S}_I \big)~,
\eea
compare with eqs. \eqref{6.3} and \eqref{6.4}.
Here we present a bi-supertwistor formulation for the superspaces 
$\overline{\mathbb M}{}^{3|2\cN} \times {\mathbb X}^\cN_m$, but first  we recall their supertwistor
realization \cite{KPT-MvU}.
  
Along with the two linearly independent even real supertwistors ${\bm T}^1 $ and $ {\bm T}^2$ 
obeying the null condition (\ref{5.3}), we also consider $m$ odd {\it complex} supertwistors
${\bm \S}^{\iu}$, with ${\iu}=1,\dots, m$, such that  (i) the bodies of ${\bm \S}^{\iu}$ are linearly independent; 
(ii) any linear combination of the supertwistors  ${\bm T}^\m $ and ${\bm \S}^{\iu}$ is null, that is
\bea
\langle {\bm T}^\m | {\bm T}^\n  \rangle_{\bm J}= \langle {\bm T}^\m | {\bm \S}^{\ju}  \rangle_{\bm J}=
\langle {\bm \S}^{\iu} | {\bm \S}^{\ju}  \rangle_{\bm J}=0~.
\eea
The supertwistors  ${\bm T}^\m $ and ${\bm \S}^{\iu}$ are  
defined modulo the equivalence  relation
\bea
({\bm T}^\m ,  {\bm \S}^{\iu}) \sim   ( {\bm T}^\n  , {\bm \S}^{\ju} )\,
\left(
\begin{array}{c||c}
 R_\n{}^\m  &B_\n{}^{\iu}  \\ \hline \hline 
0 &D_{\ju}{}^{\iu}
\end{array}
\right) ~,\quad 
\left(
\begin{array}{c||c}
 R  &B  \\  \hline \hline
 0 & D
\end{array}
\right) \in \sGL(2|m, {\mathbb C})~, \quad  R \in \sGL(2,{\mathbb R})~.~~~~~
\eea
We emphasize that the fermionic $ B_\n{}^{\iu} $ and bosonic $ D_{\ju}{}^{\iu}$ 
matrix elements  are  complex.
The space $\overline{\mathbb M}{}^{3|2\cN} \times {\mathbb X}^\cN_m$ is defined to consist 
of the equivalence classes associated with all possible  $({\bm T}^\m ,  {\bm \S}^{\iu}) $ 
under the above conditions.

It is necessary to point out two important features of the construction under consideration. 
Firstly, the invariant inner product $\langle ~, ~ \rangle_{\bm J}$ 
possesses the  property 
\bea
\langle {\bm \cT}_1 | {\bm \cT}_2  \rangle_{\bm J} 
= -(-1)^{\e_1 \e_2} \langle {\bm \cT}_2 | {\bm \cT}_1  \rangle_{\bm J} ~,
\eea
where $\e_{1,2}$ denotes the Grassmann parity of ${\bm \cT}_{1,2}$.
Secondly, associated with the odd supertwistors  ${\bm \S}^{\iu} = ( {\bm \S}_A{\,}^{\iu}) $ 
are their complex conjugates
 $\bar{\bm \S}^{\iu}= ( \bar{\bm \S}_A{\,}^{\iu})$ which possess analogous properties 
 \bea
\langle {\bm T}^\m | \bar{\bm \S}^{\ju}  \rangle_{\bm J}=
\langle \bar{\bm \S}^{\iu} | \bar{\bm \S}^{\ju}  \rangle_{\bm J}=0~.
\eea
It can be seen that the $2m$ supertwistors  ${\bm \S}^{\iu}$ and $\bar{\bm \S}^{\ju}$
are linearly independent,
\bea
\det \,\langle {\bm \S}^{\iu} | \bar{\bm \S}^{\ju}  \rangle_{\bm J}\neq 0~.
\eea

We are prepared to introduce 
a bi-supertwistor realization for $\overline{\mathbb M}{}^{3|2\cN} \times {\mathbb X}^\cN_m$.
It is given in terms of pairs $({\bm X}_{AB} , {\bm \S}_A{}^{\iu} )$, 
where the bi-supertwistor ${\bm X}_{AB}$ obeys the conditions formulated in 
subsection \ref{subsection6.2}. As to the odd supertwistors $ {\bm \S}^{\iu} $, they must be such that 
(i) the body of the $\cN \times m$ matrix  ${\bm \S}_{I}{}^{\ju} $ has rank $m$; and (ii) 
the null conditions hold
\bea
(-1)^{\e_B}{\bm X}_{AB}  {\bm J}^{BC}{\bm \S}_C{}^{\iu} 
= {\bm \S}_B{}^{\iu} {\bm J}^{BC} {\bm X}_{CA}
=0~, \qquad
{\bm \S}_{A}{}^{\iu}  {\bm J}^{AB}{\bm \S}_B{}^{\ju} =0~.
\eea
The superspace $\overline{\mathbb M}{}^{3|2\cN} \times {\mathbb X}^\cN_m$ is obtained by factorizing 
the space of all such pairs $({\bm X}_{AB} , {\bm \S}_A{}^{\iu} )$ with respect to the equivalence relation
\begin{subequations}
\bea
({\bm X}_{AB} , {\bm \S}_A{}^{\iu} ) ~\sim ~ ( \tilde{\bm X}_{AB} , \tilde{\bm \S}_A{}^{\iu} ) ~,
\eea
where 
\bea
\tilde{\bm X}_{AB}  &=& r\, {\bm X}_{AB} ~, \qquad r \in {\mathbb R}-\{0\}~, \label{7.10a}\\
\tilde{\bm \S}_A{}^{\iu} &=&(-1)^{\e_B}{\bm X}_{AB}  {\bm J}^{BC}{\bm \X}_C{}^{\iu} 
+ {\bm \S}_A{}^{\ju}  D_{\ju}{}^{\iu}~, \qquad D \in \sGL(2,{\mathbb C})~, 
 \label{7.10b}
\eea
\end{subequations}  
where ${\bm \X}^{\iu} $ are two arbitrary odd supertwistors. 
This realization for $\overline{\mathbb M}{}^{3|2\cN} \times {\mathbb X}^\cN_m$ is clearly equivalent to the
supertwistor one.

\section{Conclusion}

The bi-supertwistor realization for  $\overline{\mathbb M}{}^{4|4\cN}$ is a natural supersymmetric 
extension of Dirac's projective lightcone construction for compactified Minkowski space. 
It was Ferrara \cite{Ferrara} who posed the problem of developing such a supersymmetric extension
back in 1974. The problem has finally been solved. The supertwistor realization for  
$\overline{\mathbb M}{}^{4|4\cN}$ can be viewed as a square root of the bi-supertwistor one. 

As shown  in  \cite{GSS} (see also \cite{Siegel2012})
the bi-supertwistor realization for  $\overline{\mathbb M}{}^{4|4}$
allows one to derive compact expressions for certain correlation functions
in $\cN=1$ superconformal field theories. 
The adequate superspace setting for $\cN=2$ supersymmetric theories 
is known to be not the conventional Minkowski superspace  ${\mathbb R}{}^{4|8}$, but rather
its harmonic/projective extension ${\mathbb R}{}^{4|8} \times {\mathbb C}P^1$. 
We believe that the  bi-supertwistor realization for 
$\overline{\mathbb M}{}^{4|8} \times {\mathbb C}P^1$ proposed in section 4
will be useful for (i) the calculation of correlation functions in $\cN=2$ superconformal field theories; 
and (ii) the construction of a manifestly $\sSU(2,2|2)$ invariant formulation for  
$\cN=2$ superconformal field theories (compare, e.g.,  with non-supersymmetric approaches \cite{CO,PV}).  
The  superconformal  Fourier expansion  \eqref{4.11}  is expected to be especially important in this context. 
 \\

\noindent
{\bf Acknowledgements:}\\
The author is grateful to Daniel Butter for useful discussions and for reading the manuscript. 
This work is supported in part by the Australian Research Council.

\appendix

\section{Spinors in 4 + 2 dimensions}\label{app_A}

In this appendix we collect the salient information about spinors in $4+2$ dimensions. 
A similar discussion can be found, e.g., in \cite{Todorov}.
 
The gamma matrices in  $4+2$ dimensions, $\G_{\ha}$,  obey the anti-commutation relations 
\bea
\{ \G_{\ha} , \G_{\hb} \} = -2 \eta_{\ha \hb} {\mathbbm 1}_8 ~, \qquad 
 \eta_{\ha \hb} ={\rm diag} (-1, +1, +1,+1,+1,-1)
 \label{A1}
\eea
and can be chosen to look like 
\bea
\G_{\ha} =\left(
\begin{array}{cc}
0  ~& \S_{\ha}\\
 \tilde{\S}_{\ha}  ~&    0
\end{array}
\right) ~,
\eea
where the $4\times 4$ matrices $\S_{\ha}$ and $ \tilde{\S}_{\ha}  $ have the explicit form 
\begin{subequations}\label{A.3}
\bea 
\S_{\ha} &=& (\S_a, \S_5, \S_6) = ( \ri \g_a , \g_5, {\mathbbm 1}_4) 
\equiv(\S_{\ha})_{\hal \, \hat \beu}
~, \\
\tilde{\S}_{\ha} &=& (\tilde{\S}_a, \tilde{\S}_5, \tilde{\S}_6) = ( -\ri \g_a , - \g_5, {\mathbbm 1}_4) 
\equiv (\tilde{\S}_{\ha})^{\hat \alu \, \hbe}
~.~~~~
\eea
\end{subequations}
Here $\g_a$ are the gamma matrices in $3+1$ dimensions,  and $\g_5 := -\ri \g_0 \g_1 \g_2 \g_3$. 
Our choice of  $\g_a$ coincides with that adopted in \cite{WB,BK}, specifically 
\bea
\g_{a} =\left(
\begin{array}{cc}
0  ~& \s_{a}\\
 \tilde{\s}_{a}  ~&    0
\end{array}
\right) 
~,\qquad
\g_{5} =\left(
\begin{array}{cc}
{\mathbbm 1}_2  ~& 0 \\
 0  ~&    -{\mathbbm 1}_2
\end{array}
\right) ~,
\eea
where 
\bea
\s_a = ({\mathbbm 1}_2 , \vec{\s}) \equiv ( \s_a)_{\a \dt \a}~, 
\qquad \tilde{\s}_a = ({\mathbbm 1}_2 ,-  \vec{\s}) \equiv ( \tilde{\s}_a)^{\dt \a \a}~.
\eea
The matrices \eqref{A.3} obey the relations 
\bea
\S_{\ha} \tilde{\S}_{\hb} +\S_{\hb} \tilde{\S}_{\ha} = -2\eta_{\ha \hb} {\mathbbm 1}_4~, 
\qquad 
\tilde{\S}_{\ha} {\S}_{\hb} + \tilde{\S}_{\hb} {\S}_{\ha} = -2\eta_{\ha \hb} {\mathbbm 1}_4~.
\label{A.6}
\eea
For the matrix $\G_7 := -\ri \G_0\G_1 \G_2\G_3\G_5\G_6$ we obtain
\bea
\G_{7} =\left(
\begin{array}{cc}
{\mathbbm 1}_4  ~& 0 \\
 0  ~&    -{\mathbbm 1}_4
\end{array}
\right) ~.
\eea

The Dirac spinor representation of the double covering group of $\sSO_0(4,2)$ is generated by 
\bea
{\frak J}_{\ha \hb } := -\frac{1}{4}[\G_{\ha} , \G_{\hb} ]  =\left(
\begin{array}{cc}
\S_{\ha \hb}   ~& 0 \\
 0  ~&    \tilde{\S}_{\ha \hb} 
\end{array}
\right) ~,
\eea
where 
\begin{subequations} 
\bea
\S_{\ha \hb} &:=& -\frac{1}{4} ( \S_{\ha} \tilde{\S}_{\hb} -\S_{\hb} \tilde{\S}_{\ha} )
\equiv (\S_{\ha \hb} )_{\hal \,}{}^{\hbe}~, \\
\tilde{\S}_{\ha\hb} &:=& -\frac{1}{4} ( \tilde{\S}_{\ha} {\S}_{\hb} - \tilde{\S}_{\hb} {\S}_{\ha} )
\equiv ( \tilde{\S}_{\ha \hb} )^{\hat  \alu \,}{}_{\hat \beu}~.
\eea
\end{subequations}
The matrices $\S_{\ha \hb} $ are the generators of the group $\sSU(2,2)$ defined by \eqref{SU(2,2)}.
The following isomorphism holds:
$\sSO_0(4,2) \cong \sSU(2,2)/{\mathbb Z}_2$. 

The Hermitian conjugation properties of the gamma matrices  are 
\bea
(\G_{\ha})^\dagger = \G_0 \G_6\,\G_{\ha}\,  \G_0 \G_6 ~,
\eea
hence 
\bea
({\frak J}_{\ha \hb})^\dagger = \G_0 \G_6 \, {\frak J}_{\ha \hb} \, \G_0 \G_6 ~,
\eea
This implies 
\bea
(\S_{\ha})^\dagger = \g_0 \tilde{\S}_{\ha} \g_0~, \qquad
(\tilde{\S}_{\ha})^\dagger = \g_0 {\S}_{\ha} \g_0~, 
\eea
and hence
\bea
(\S_{\ha \hb})^\dagger = - \g_0 {\S}_{\ha \hb} \g_0~, \qquad
(\tilde{\S}_{\ha \hb})^\dagger = \g_0 {\tilde{\S}}_{\ha \hb} \g_0~. 
\eea
It  can be seen that $\g_0$ coincides with the matrix $\O$ in eqs. \eqref{2.6} and \eqref{SU(2,2)}.

Given a Dirac spinor 
\bea
{\bm \J}  =\left(
\begin{array}{c}
\j \\
\f
\end{array} 
\right) ~, \qquad \j = (\j_{\hal})~, \quad
\f = ( \f^{\hat \alu} )~.
\eea
its Dirac conjugate is defined as  follows:
\bea
\overline{\bm \J} := -\ri \,{\bm \J}^\dagger \G_0\G_6 = (\j^\dagger \g_0 , - \f^\dagger \g_0) ~, 
\qquad \j^\dagger \g_0 \equiv (\bar \j^{\hal}) ~, \quad
\f^\dagger \g_0 \equiv (\bar \f_{\hat \alu})~.
\eea
The infinitesimal $\sSO(4,2)$ transformation laws of these spinors are:
\begin{subequations}
\bea
\d {\bf \J} &=&\phantom{-} \hf \o^{\ha \hb} {\frak J}_{\ha \hb} {\bf \J}~, \\
\d \overline{\bf \J} &=& - \hf \overline{\bf \J}
\o^{\ha \hb} {\frak J}_{\ha \hb} ~.
\eea
\end{subequations}
The Dirac spinor representation is a sum of two irreducible ones, 
one of which is the twistor representation and the second is equivalent to its dual (contragredient).
The twistor representation is associated with spinors of the form 
\bea
{\bm \J}_{\rm L}  =\left(
\begin{array}{c}
\j \\
0
\end{array} 
\right) ~, \qquad \j = (\j_{\hal})~
\eea
such that $\G_7 {\bm \J}_{\rm L}  ={\bm \J}_{\rm L}  $.
The Dirac conjugate of ${\bm \J}_{\rm L}  $, 
\bea
\overline{\bm \J}_{\rm L} = (\bar \j , 0)~, \qquad \bar \j := \j^\dagger \g_0 = (\bar \j^{\hal})
\eea
transforms according to the dual twistor representation. 
The infinitesimal $\sSO(4,2)$ transformation laws of ${\bm \J}_{\rm L} $ and  
$\overline{\bm \J}_{\rm L} $ are:
\begin{subequations}
\bea
\d \j_{\hal}  &=& \hf \o^{\hc \hd} (\S_{\hc \hd})_{\hal}{}^{\hbe} \j_{\hbe}~, \\
\d \bar \j^{\hal}  &=& - \hf  \bar \j^{\hbe}  \o^{\hc \hd} (\S_{\hc \hd})_{\hbe}{}^{\hal} ~.
\eea
\end{subequations}
Explicit calculations give
\bea
\hf \o^{\ha \hb} \S_{\ha \hb} =\left(
\begin{array}{c|c}
  \hf \o^{ab} \s_{ab} - \t {\mathbbm 1}_2\phantom{\Big|}& -\ri b^a \s_a \\
 \hline
-\ri a^a \tilde{\s}_a   &\phantom{\Big|}      \hf {\o}^{ab} \tilde{\s}_{ab} + \t {\mathbbm 1}_2
\end{array}
\right) ~,
\eea
where the parameters 
$\t = \hf \o^{56}$, $a^a = \hf (\o^{a6} - \o^{a5}) $ and $b^a = \hf (\o^{a6} + \o^{a5}) $ 
generate a dilatation, a space-time transaction and a special conformal transformation respectively. 
As usual, the $2\times 2$ matrices ${\s}_{ab} $ and $\tilde{\s}_{ab} $ denote the Lorentz generators 
of the (1/2, 0) and (0,1/2) representations of the Lorentz group in four dimensions \cite{BK}, 
\bea
\s_{a b} := -\frac{1}{4} ( \s_{a} \tilde{\s}_{b} -\s_{b} \tilde{\s}_{a} )~,
\qquad
\tilde{\s}_{a b} := -\frac{1}{4} ( \tilde{\s}_{a} {\s}_{b} - \tilde{\s}_{b} {\s}_{a} )~.
\eea

Since the matrices $\G_{\ha}$ and $- \G_{\ha}^{\rm T}$ constitute two representations 
of the same Clifford algebra, eq. \eqref{A1},  and these representations are necessarily equivalent, 
we have 
\bea
{\frak C}^{-1} \G_{\hat a} {\frak C} = - \G_{\ha}^{\rm T} ~,
\eea
hence
\bea
{\frak C}^{-1} {\frak J}_{\hat a \hb} {\frak C} = - {\frak J}_{\ha \hb}^{\rm T} ~,
\eea
for some charge conjugation matrix $\frak C$. It can be chosen as 
\bea
{\frak C} = \left(
\begin{array}{cc}
0   ~& \g_5 C  \\
 -\g_5 C ~&  0
\end{array}
\right) 
\equiv \left(
\begin{array}{cc}
   0 ~& {\frak C}_{\hal}{\,}^{\hat \beu}  \\
 {\frak C}^{\hat \alu}{\,}_{\hbe}   ~&   0
\end{array}
\right) ~,
\eea
with $C$ the  charge conjugation matrix in $3+1$ dimensions, which is defined by 
\bea
C^{-1}\g_a C = - \g_a^{\rm T}\quad \longrightarrow \quad C^{-1}\g_5 C =  \g_5^{\rm T}
\eea
and can be chosen as 
\bea
C= \left(
\begin{array}{cc}
\ve_{\a\b}    ~& 0  \\
 0 ~&  \ve^{\dt \a \dt \b}
\end{array}
\right) ~, \qquad 
C^{-1}= \left(
\begin{array}{cc}
\ve^{\a\b}    ~& 0  \\
 0 ~&  \ve_{\dt \a \dt \b}
\end{array}
\right) ~.
\eea
Using the properties  $C^\dagger = C^{\rm T} = -C = C^{-1}$
gives 
\bea
{\frak C}^{\rm T} =  {\frak C} ~.
\eea
The inverse of $\frak C$ is 
\bea
{\frak C}^{-1} = \left(
\begin{array}{cc}
   0 ~& ({\frak C}^{-1})^{\hal}{\,}_{\hat \beu}  \\
( {\frak C}^{-1})_{\hat \alu}{\,}^{\hbe}   ~&   0
\end{array}
\right) ~.
\eea
Since $\frak C$ is symmetric, it holds that 
\bea
 {\frak C}_{\hal}{\,}^{\hat \beu}  =  {\frak C}^{\hat \beu}{\,}_{\hal} ~, \qquad
 ({\frak C}^{-1})^{\hal}{\,}_{\hat \beu}  = ( {\frak C}^{-1})_{\hat \beu}{\,}^{\hal} ~.
\label{A.22}
\eea
Given a Dirac spinor $\bm \J$, its charge conjugate spinor defined by 
\bea
{\bm \J}_{\frak C} := {\frak C}  \overline{\bf \J}^{\rm T}
\eea
transforms as  a Dirac spinor, 
\bea
\d {\bf \J}_{\frak C} = \hf \o^{\ha \hb} {\frak J}_{\ha \hb} {\bf \J}_{\frak C}~.
\eea

The transformation laws of $\bm \J$, $\overline{\bf \J} $ and ${\bm \J}_{\frak C} $ show that 
\bea
\f^{\hal} :=  ({\frak C}^{-1})^{\hal}{\,}_{\hat \beu} \f^{\hat \beu}
=  \f^{\hat \beu} ({\frak C}^{-1})_{\hat \beu}{\,}^{ \hal} 
\eea
transforms as a dual twistor, while 
\bea
\bar \f_{\hal} :=  {\frak C}_{\hal}{\,}^{\hat \beu} \bar \f_{\hat \beu}
=  \bar \f_{\hat \beu} {\frak C}^{\hat \beu}{\,}_{\hal}
\eea
transforms as a twistor. 
This means that the matrices  \eqref{A.22} are invariant tensors of $\sSU(2,2)$ which 
can be used to convert all underlined twistor indices, $\hat \alu, \hat \beu, \dots,$
into  twistor indices, $ \hal,  \hbe, \dots$, and therefore to completely get rid of the former.

The sigma matrices with twistor indices are 
\bea
(\S_{\ha})_{\hal \hbe} := (\S_{\ha})_{\hal \hat \gau}  {\frak C}^{\hat \gau}{\,}_{\hal}~, \qquad
(\tilde{\S}_{\ha})^{\hal \hbe} := ({\frak C}^{-1})^{\hal}{\,}_{\hat \gau}
(\S_{\ha})^{\hat \gau \hbe}~.
\label{A.32}
\eea
Since $(\G_{\hat a} {\frak C})^{\rm T} = - \G_{\hat a} {\frak C}$, these matrices are antisymmetric, 
\bea
(\S_{\ha})_{\hal \hbe} = -(\S_{\ha})_{ \hbe  \hal} ~, \qquad
(\tilde{\S}_{\ha})^{\hal \hbe} = - (\tilde{\S}_{\ha})^{ \hbe \hal }~.
\eea
The matrices $(\S_{\ha})_{\hal \hbe} $ and $(\tilde{\S}_{\ha})^{\hal \hbe}$ obey the relations \eqref{A.6}.
The following completeness relations hold:
\begin{subequations}\label{A.14}
\bea
\hf \ve_{\hal \hbe \hga \hde}  (\tilde{\S}^{\ha})^{\hga \hde}&=& ({\S}_{\ha})_{\hal \hbe}~,\\
 \hf \ve^{\hal \hbe \hga \hde}  (\S_{\ha})_{\hga \hde}&=& (\tilde{\S}^{\ha})^{\hal \hbe}~,\\
 \hf (\S^{\ha})_{\hal \hbe} (\S_{\ha})_{\hga \hde} &=&
\ve_{\hal \hbe \hga \hde} 
~.
\eea
\end{subequations}

There is a one-to-one correspondence between complex vectors in $4+2$ dimensions 
$V^{\ha}$ and bi-twistors $V_{\hal \hbe} = - V_{\hbe \hal}$ or 
 $V^{\hal \hbe} = \hf  \ve^{\hal \hbe \hga \hde}  V_{\hga \hde}$.
It is described by the relations 
\begin{subequations}
\bea
V_{\hal \hbe} &=& V^{\hc } (\S_{\hc} )_{\hal \hbe} ~, \qquad 
V^{\ha} = \frac{1}{4}(\tilde{\S}^{\ha})^{ \hga \hde}  V_{\hga \hde} ~, \\
V^{\hal \hbe} &=& V^{\hc } (\tilde{\S}^{\hc} )_{\hal \hbe} ~, \qquad 
V^{\ha} = \frac{1}{4}({\S}_{\ha})^{ \hga \hde}  V^{\hga \hde} ~.
\eea
\end{subequations}
It is convenient to use the matrix notation $V:= (V_{\hal \hbe} )$ and $\widetilde{V}:= (V^{\hal \hbe} )$.
If $\bar V^{\ha}$ is the  complex conjugate of  $V^{\ha}$, then the corresponding bi-twistor matrices 
$\widetilde{\bar V}:= (\bar V^{\hal \hbe} )$ and $V:= (V_{\hal \hbe} )$ are related to each other as in 
\eqref{17conj}.
Finally, if $V^{\ha}$  is real, then $\widetilde{\bar V}= \widetilde{ V}$.

\begin{footnotesize}

\end{footnotesize}

\end{document}